\documentclass{aa}

\usepackage{graphicx}
\usepackage{natbib}
\usepackage{txfonts}
\usepackage{amsmath}
\begin{document} 

\title{Scattering linear polarization of late-type active stars}

\author{T.M. Yakobchuk
        \and
        S.V. Berdyugina}

\institute{Kiepenheuer-Institute f\"{u}r Sonnenphysik, 
           Sch\"{o}neckstr.6, D-79104, Freiburg, Germany\\
\email{yakobchuk@leibniz-kis.de, sveta@leibniz-kis.de}}

\date{Received -; accepted -}

 
  \abstract
   {Many active stars are covered in spots, much more so
   than the Sun, as indicated by spectroscopic and photometric observations.
   It has been predicted that star spots induce non-zero intrinsic linear polarization by breaking the visible stellar disk symmetry.
   Although small, this effect might be useful for star spot
   studies, and it is particularly significant for a future polarimetric atmosphere
   characterization of exoplanets orbiting active host stars.}
   {Using models for a center-to-limb variation of the intensity and polarization
   in presence of continuum scattering and adopting a simplified two-temperature
   photosphere model, we aim to estimate
   the intrinsic linear polarization for late-type stars of different gravity,
   effective temperature, and spottedness.}
   {We developed a code that simulates various spot configurations or uses
   arbitrary surface maps, performs numerical disk integration, and builds
   Stokes parameter phase curves for a star over a rotation period for a
   selected wavelength.
   It allows estimating minimum and maximum polarization values
   for a given set of stellar parameters and spot coverages.}
   {Based on assumptions about photosphere-to-spot temperature contrasts and spot size
   distributions, we calculate the linear polarization
   for late-type stars with $T_{\rm eff}$ = 3500~K - 6000~K, log$g$ = 1.0 - 5.0,
   using the plane-parallel and spherical atmosphere models. Employing random spot
   surface distribution, we analyze the relation between spot coverage and polarization
   and determine the influence of different input parameters on results. Furthermore,
   we consider spot configurations with polar spots and active latitudes and
   longitudes.}
   {}

   \keywords{starspots -- stars: polarization -- stars: activity}

   \maketitle
%

\section{Introduction}

Numerous studies on stellar activity have been performed in the past decades using spectroscopic
Doppler imaging and photometric light curve inversion techniques 
\citep[e.g., see reviews by][]{2005LRSP....2....8B, 2009A&ARv..17..251S}.
Together with the prospects
of using polarimetry to characterize exoplanet atmospheres
\citep[e.g.,][]{2000ApJ...540..504S, 2008ApJ...673L..83B}, this
raises the interest in stellar polarimetry. The field is also favored by current instrumental progress because
detecting stellar polarization often demands high accuracies of up to 10$^{-5}$-10$^{-6}$ .
In particular, two major surveys by \cite{2010MNRAS.405.2570B} and \cite{2016MNRAS.455.1607C}
have been completed in recent years, totaling nearly 100 nearby bright stars measured
with polarization
sensitivities at the parts-per-million level. Most targets from the sample of Northern bright 
stars by \cite{2010MNRAS.405.2570B} showed measurable linear polarization that was mainly of
interstellar origin, however, and only for a few stars was the
polarization deduced to be due to intrinsic polarigenic
mechanisms. The subsequent study of \cite{2016MNRAS.455.1607C} for Southern hemisphere stars found
stronger polarizations in general, with an intrinsic mechanism identified for more than 20 stars.

Considering the different spectral types, evolutionary status, environment, rotation,
and binarity, the intrinsic stellar polarization can be explained by different mechanisms.
However, the main sources of polarization are usually related to either light scattering or
magnetic fields. Scattering by free electrons or dust (in cooler environments)
results in the polarization of stars or binaries that have circumstellar asymmetric envelopes
\citep[e.g.,][]{1977A&A....57..141B, 1979ApJ...232..181P}, while the so-called
magnetic intensification predicts
the net linear polarization from optically thick spectral lines in strong magnetic
fields \citep[e.g.,][]{1962AnAp...25..127L, 1976ApJ...204..818M}. In addition to
circumstellar scattering, \cite{1969ApL.....3..165H} proposed another mechanism:
temperature variations in the photosphere that produce a disk asymmetry might also lead to 
the observable net scattering polarization.

Early observations by \cite{1971AJ.....76..431D}, who performed
a $UBV$ broadband polarimetric
survey of 55 cool stars of K and
M type, showed intrinsic polarization for late-type supergiants and no polarization
for class III giants. When the survey was complemented with infrared (IR) photometry
\citep{1971ApJ...165...57D} and near-IR polarimetry \citep{1971AJ.....76..901D}, the
polarization was associated with the optically thin circumstellar clouds of solid particles.
\cite{1977A&AS...30..213P} and \cite{1982A&A...105...53T}
investigated the interstellar polarization in the solar neighborhood and concluded that
any intrinsic polarization of the stars in the sample, if present, is close to the error limits,
which precluded them from drawing further conclusions. However, they indicated that there is a tendency for
higher polarizations to be mostly confined to the stars with spectral type F and later.
\cite{1985A&A...152..357H} reported five-color ($UBVRI$) polarimetric observations of 13 solar-type
stars that showed a clear wavelength dependence. This decreased from ultraviolet to red,
with time variations. The observed linear polarization was suggested to be at least partly
intrinsic and was explained by either magnetic intensification
or scattering in the presence of
spot-induced disk asymmetry. \cite{1989A&AS...78..129H} described
the magnetic intensification as the dominant effect
in the linear polarization in the most active single late-type dwarfs (F7$-$K5),
while based on the wavelength dependence,
it might result from a combination of Rayleigh and Thompson scattering
in less active
dwarfs.

In the recent study by \cite{2016MNRAS.455.1607C}, most stars in the sample showed
positive detections with polarizations of more than $10^{-5}$, with several classical Be stars
exceeding $10^{-3}$. Significant intrinsic polarization was found in most B stars in the sample,
and it was also apparent for a number of late-type giants.
It was attributed mainly to circumstellar dust shells. However, in some
cases, other explanations were used. For the dust-free M giant $\delta$ Oph, for instance,
the variable and high-degree polarization was associated with Rayleigh scattering in the
photosphere. 
Another recent study by \cite{2017MNRAS.467..873C} focused on nearby FGK dwarfs, for which they measured the
linear polarization of 32 active, inactive, and debris-disk host stars with the same high
precision. It was found that active stars in the sample have a mean polarization of
(23.0 $\pm$ 2.2) $\times$ 10$^{-6}$ (after subtracting the interstellar polarization),
with maximum levels ranging up to $\sim$45 $\times$ 10$^{-6}$. In their discussion of the mechanisms,
the authors suggested the magnetic intensification as a most likely mechanism for active late-type
dwarfs.

Binary stars reveal polarization that in most cases is produced by the same phenomena as in
single stars, but it involves a more complex geometry and new asymmetries  
\citep[see the review by][]{2005ASPC..343..389M}. One particular class of
non-eclipsing active binaries that are frequent in polarimetric studies
is RS CVn-type close binaries, where one component is usually an active late-type evolved
star and the second component is a faint unresolvable dwarf.
\cite{1993A&AS..102..343S} measured $UBVRI$ linear and circular
polarization in 15 RS CVn-type
binaries, of which 6 systems showed evidence of circumbinary dust envelopes from the IR excess,
6 binaries did not show the excess, and the rest had uncertain detections. Later polarimetric
observations of RS CVn systems reached similar conclusions, ascribing detected signals either to
circumbinary envelopes or to the interstellar medium
\citep[e.g.,][]{2002A&A...386..916Y,2006ApJ...651..475B,2009MNRAS.396.1004P}.

In this paper, we simulate the intrinsic linear polarization of late-type active stars that is due
to spot-induced disk asymmetry. The simulations are based on the calculated data for the scattering polarization
in continuum for plane-parallel and spherical stellar models. In Section 2 we describe the method we developed
to generate spots of different size and surface distribution, and perform the disk
integration to obtain the total Stokes parameters. In Section 3 we
place constraints on the maximum polarization levels for different stellar models and
present results for a random spot surface distribution. We analyze how our results
depend on the inclination, wavelength, and photosphere-to-spot temperature contrast.
In a few subsections, we consider the polarization signatures expected from polar spots and active latitudes and
longitudes. Section 4 concludes the paper.

\section{Method}

\subsection{Generating spots}

We developed a code that generates different spot distributions by coordinates, size,
and temperature to calculate light curves and Stokes parameter variations over a
stellar rotation period. Previously obtained surface maps and
Doppler-imaging temperature/filling factor maps can also be used
as input. The stellar surface is
represented by a uniform spherical grid over longitude and latitude.
For our simulations we chose a 1$^{\circ} \times$1$^{\circ}$ grid resolution, which
gives optimal accuracy and short computation times. The main model ingredients were
chosen as described below.

 \subsubsection{Temperatures}
 Different methods indicate a tendency for
 hotter stars to have more highly contrasting spots than cooler stars, which seems to hold
 for active stars independent of their luminosity class \citep{2005LRSP....2....8B}.
 Based on available published data (which are still scarce and inhomogeneous),
 \cite{2015MNRAS.448.3053A} (see the dashed blue line in Fig.
2) used a linear fit to describe
 the dependence between the effective temperature $T_{\rm eff}$ of a star and the spot temperature
 contrast $\Delta T$. We adopted it in our simulations to determine
the
 temperature of spot umbrae as follows:
 
  \begin{equation} 
    \Delta T = 0.45 T_{\rm eff} - 1080.
   \label{eq_dt}
  \end{equation}
 
 By setting the corresponding area and temperature ratios, we also added an
 option to include the penumbra for each spot. It was selected in our calculations,
 with the temperature ratio for penumbrae set to the mid-point between the photosphere and
 the umbra.

 \subsubsection{Sizes}
 With the exception of the Sun, no information is available about
 the spot size distribution on other stars, not to mention its dependence on the activity
 and spectral type. Therefore, spot sizes in the program can be set either manually 
 (along with the coordinates), uniformly random within the specified size range,
 or following normal or lognormal distribution.
 
 In the first statistically significant study by
 \cite{1988ApJ...327..451B}, which  covered the period from 1917 to 1982, sunspot umbral areas
 were found to be distributed lognormally. Using parameters of the lognormal fit from
 this study, obtained for combined solar minimum and maximum, \cite{2004MNRAS.348..307S} 
 extrapolated them to higher activity levels of solar-like stars, suggesting several scaling
 laws. \cite{1988ApJ...327..451B} gave the size distribution as
 \begin{equation}
  \frac{dN}{dA} =  \left( \frac{dN}{dA} \right)_{max}
                   \exp \left[ - \frac{(\ln A - \ln \langle A \rangle)^2}{2 \ln \sigma_A} \right] ,
  \label{eql0}
 \end{equation}
 
 \noindent where $\langle A \rangle$ is the mean spot umbral area, and $\sigma_A$
 determines the width of the lognormal distribution. \cite{1988ApJ...327..451B} expressed
 the area and the width of the distribution in units of $10^{-6} A_{\rm 1/2 \odot}$. 
 \cite{2004MNRAS.348..307S} suggested
 that these parameters change with the total starspot area coverage as

 \begin{equation}
   \sigma_A = \sigma _A ^0 + \Delta _\sigma (A_{\rm spot}/A_*) ^{n_\sigma}
  \label{eql1}
 \end{equation}

 \begin{equation} 
   \langle A \rangle = \langle A \rangle ^0 + \Delta _A (A_{\rm spot}/A_*) ^{n_A} ,
  \label{eql2}
 \end{equation}
 
 where $\Delta _\sigma$ and $\Delta _A$ are coefficients determining the increase in the
 width and mean area of the spot size distribution with increasing activity, expressed as the ratio
 of the area occupied by spots to the total stellar surface area $A_{\rm spot}/A_*$. Next, 
 $\sigma _A ^0$ and $\langle A \rangle ^0$ are parameter values in the limit of zero spot coverage.
 It is assumed that the dependence on the activity is a power
law, with exponents $n_\sigma$ and $n_A$.
 In practice, $\Delta _\sigma$ and $\Delta _A$
 and the zero-limit parameter values were determined by solving Eqs. \ref{eql1} and \ref{eql2}
 using the known spot coverage and the distribution width of the solar minimum and maximum
 from \cite{1988ApJ...327..451B} for a selected power $n_\sigma$ or $n_A$.
 
 It should be noted that except for the paper by \cite{2004MNRAS.348..307S} for solar-like stars,
no comprehensive studies have been devoted to spot size distributions in stars of different
 spectral type and activity \citep[for low-mass stars, also see][]{2013MNRAS.431.1883J}.
 Current Doppler-imaging studies, which potentially  approach the topic most directly, offer a resolution no better than several square degrees
 \citep[e.g.,][]{2009A&ARv..17..251S, 2016LNP...914..177K},
 which is not sufficient for meaningful size distribution constraints. 
 Although Doppler maps of many stars reveal large
 active areas, it is therefore not possible to distinguish whether they are homogeneous structures or
 agglomerates of many smaller spots.
 
 Because \cite{2004MNRAS.348..307S} only extrapolated their size distributions for
 solar-like stars,
 additional assumptions were needed for other late-type stars. By simply using the scaling laws
 with solar values  for other stars, we obtain overly small spots for giants and too
 large spots for dwarfs, both of which seems improbable. We finally
adopted the scaling relations
 \ref{eql1} and \ref{eql2}, but instead of absolute solar spot sizes, we chose relative
 angular sizes with the same parameter values, but in stellar surface area units of $10^{-6} A_{\rm 1/2 *}$.
 
 Using the scaling law in Eq. \ref{eql1} for
 $\sigma_A$, we assumed that the other parameter,
 $\langle A \rangle$, was fixed at $0.57 \times 10^{-6} A_{\rm 1/2 *}$
 \citep[i.e., equal to the solar value at $R_*$ = 1 $R_\odot$ from][]{1988ApJ...327..451B}. Accordingly, when
 scaling $\langle A \rangle$ with Eq. \ref{eql2}, the distribution
width,
 $\sigma_A$, remained
 constant and equal to $4 \times 10^{-6} A_{\rm 1/2 *}$. Owing to the ambiguity of the results from
 varying two parameters simultaneously, \cite{2004MNRAS.348..307S} refrained from considering these
 extrapolations in detail. Below we present the results we obtained using scalings for $\sigma_A$
 and $\langle A \rangle$ for selected ${n_\sigma}$ and ${n_A}$ exponents in the range between
 0.5 (square-root law) to 1.0 (linear), following the prescriptions
of \cite{2004MNRAS.348..307S}.

 Finally, the starspots in our study were assumed to have circular shape and to consist of umbral
 and penumbral regions with areas ratios 1:4. Similarly to \cite{2004MNRAS.348..307S},
 the area in Eq. \ref{eql0} also corresponded to the umbral area,  while for
 the total spot coverage in Eqs. \ref{eql1} and \ref{eql2}, we integrated the stellar surface grid by
 pixels including both umbral and penumbral parts.

 \subsubsection{Coordinates}
 We added several options for how spots
 could be distributed on the stellar surface. Along with the manually set coordinates, this includes
 random distribution, active latitudes and longitudes, and polar spot generation. The simulation for
a  random spot distribution implies that the spots
 are distributed uniformly randomly in longitude, $\phi$, and in sine of latitude, sin($\theta$), to
 produce an equal probability across the stellar surface. 
 We avoided overlapping of generated spots: before we placed each spot, we iteratively checked for
 overlaps with umbral regions of previously
added spots on the surface map and then relocated when necessary (1000 iterations at most).
 
Additionally, we assumed that spots do not evolve in time in size and position, and
there is no differential stellar rotation, although this can be readily implemented in the code.

\subsection{Integrating the disk}

Our calculations are based on the previously developed code for exoplanet transit
polarization simulations \citep{2015ApJ...806...97K}, and it employs two-dimensional grid integration.
Here we chose a uniform Cartesian grid representing the sky plane, integrating over only those grid
pixels whose centers are inside the stellar disk.
We tried different binning as well as a more advanced polar grid integration that accounted for
partially divided pixels (this includes both spot and photosphere-projected stellar surface elements)
to test the reliability of the method.
The uniform grid was found to be sufficiently accurate for resolutions
after a certain level and did not show any noticeable artifacts related to the pixelation of
the limb. Similarly, we found that partially divided pixels do not need special consideration.
A grid of 1000x1000 pixels size was chosen, which is both fast and accurate enough for the required
polarization levels.

During integration, we transformed the coordinates of each grid pixel into the spherical coordinate system of the star, finding the corresponding longitude/latitude and the temperature (or the filling factor), according
to a given stellar rotation axis orientation and phase angle.
The total Stokes parameters at each rotation phase were calculated by summing all grid
pixels of the visible stellar disk as follows:

\begin{equation}
F = \sum_{i,j} a_{ij}~I(\mu_{ij})
\label{eqlF}
\end{equation}

\begin{equation}
q = \sum_{i,j} a_{ij}~I(\mu_{ij}) P(\mu_{ij}) \cos2\phi_{ij}~/~F
\label{eqlQ}
\end{equation}

\begin{equation}
u = \sum_{i,j} a_{ij}~I(\mu_{ij}) P(\mu_{ij}) \sin2\phi_{ij}~/~F
\label{eqlU}
,\end{equation}

\noindent where $a$ is the area of the $i,j$ pixel (constant for a regular grid),
$\mu$ is the angle  between the surface normal and the line of sight to the observer,
$\phi$ is the polar angle of a system with the origin at the disk center,
$F$ is the total stellar flux, and $q$ and $u$ are normalized Stokes
parameters. We note that the relative flux is normalized to
the total flux of an unspotted photosphere in our analysis.
The center-to-limb variations of intensity $I(\mu_{ij})$ and polarization $P(\mu_{ij})$
were found through trilinear interpolation using the look-up tables from \cite{2015A&A...575A..89K}
and \cite{2016A&A...586A..87K}, according to the selected wavelength, surface gravity, and
temperature of a star. \cite{2015A&A...575A..89K} calculated the tables for continuum spectra
of FGK stars ($T_{\rm eff}$ = 4500~K - 6900~K, log$g$ = 2.0 - 5.0, and wavelength range 
4000 - 7000 \text{\AA}) for the Phoenix grid of 
plane-parallel models, and in \cite{2016A&A...586A..87K} similar calculations were
made for a wider range of models ($T_{\rm eff}$ = 4000~K - 7000~K
and log$g$ = 1.0 - 5.5)
assuming a spherical stellar atmosphere. In addition, we used unpublished data on
intensity and polarization variations for both plane-parallel and spherical cooler models with
temperatures down to 3000~K, obtained with the same code by N. Kostogryz
(2016, private communication). It should be noted that these data for cooler stars
do not include atomic and molecular absorption lines, which can lead to overestimated polarization values, especially for blue wavelengths and for dwarfs with higher surface
gravities (depending on the selected wavelength). Evidently, a proper
spectrum synthesis code is needed to calculate the intrinsic polarization for these cases. However, the depolarizing effect from absorption lines may not be very significant because when the absorption line suppresses the polarization, the intensity is also reduced.
In turn, this will affect the normalized polarization parameters that we calculate less strongly. Additionally,
coherent scattering processes, especially in molecular bands, as seen on the Sun
\citep[e.g.,][]{2002A&A...388.1062B} probalby increase polarization
further, which counteracts the
effect of the absorption lines.


\section{Results}

\subsection{Constraining the upper  polarization limits}

\begin{figure}
\includegraphics[height=8.5cm, angle=-90]{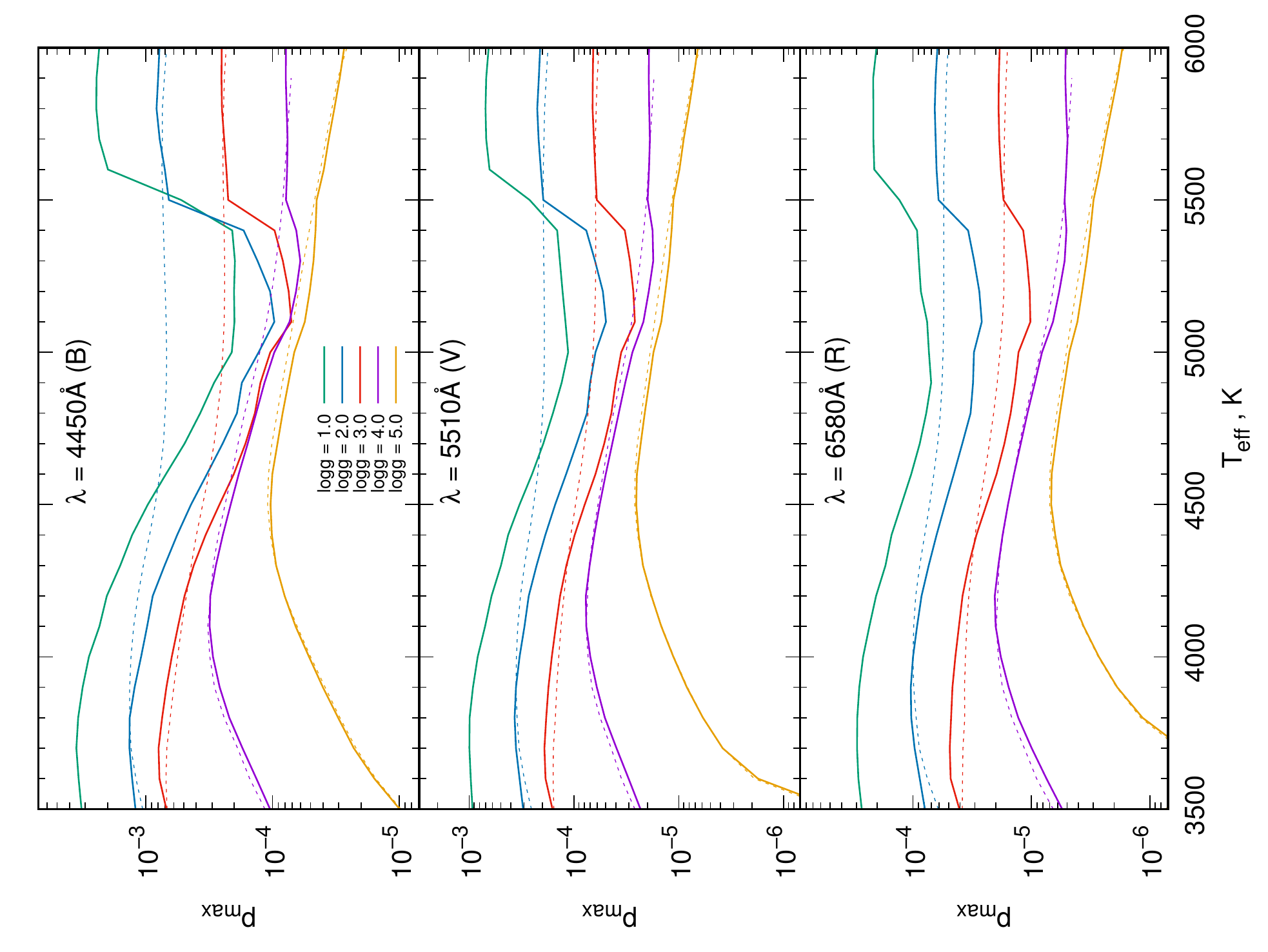}
\caption{Highest polarization degree as a function of effective temperature for
different types of stars in selected wavelengths using plane-parallel 
(\textit{dashed}) and spherical (\textit{solid})
model atmospheres.
\label{fig_pmax}}
\end{figure}

\begin{figure}
\includegraphics[width=8.8cm, angle=0]{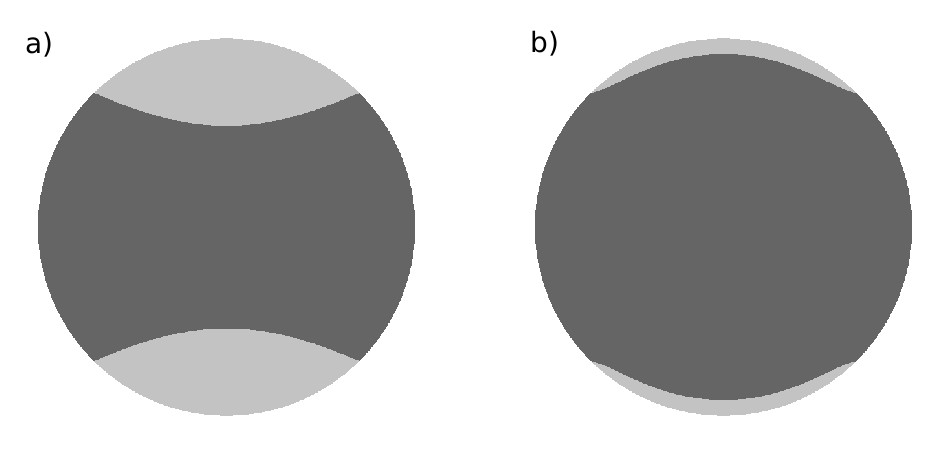}
\caption{Spotted area distribution on a stellar disk
resulting in the highest polarization for two models
at $\lambda$ = 4450 \text{\AA}:
a) $T_{\rm eff}$ = 3500~K, $T_{\rm spot}$ = 3000~K, and log$g$ = 2.0, and 
b) $T_{\rm eff}$ = 5100~K, $T_{\rm spot}$ = 3875~K, and log$g$ = 2.0.
The spotted area is shown with the darker color. For other stellar
models, the highest possible polarizations are obtained at intermediate spot
coverage with similar disk appearances, in which photosphere regions
at the top and bottom vary between the two cases.
\label{fig_pmaxdsk}}
\end{figure}

According to the definition of the Stokes $Q$ and $U$ parameters, for any arbitrarily
spotted disk, a direction exists in which $Q$ takes its maximum value and $U$
is zero. Thus, the problem of determining the highest possible polarization degree for a
given star can be reduced to the problem of determining the highest normalized $Q/I$. Since
the Stokes $Q$ is proportional to the cosine of the double angle, the stellar disk
is divided into four $Q$ quadrants of opposite sign that cancel each other out.
Every point at a distance $\mu$ from the disk center has seven counterpart points in
the other octants that can take the same $Q$ and $I$ absolute values. In the
two-temperature model, the highest total polarization
from the combination of these eight points is obtained when
$\Delta Q_{ij} ^{max} = 4 \times (Q_{ij} ^{\mathrm{phot}} - Q_{ij} ^{\mathrm{spot}})$,
with the total flux being
$4 \times (F_{ij} ^{\mathrm{phot}} + F_{ij} ^{\mathrm{spot}})$.
Searching for the normalized polarization degree, that is, for the ratio of the total
polarized flux to the total flux from a star, we began by calculating the minimum total
flux $F_{\rm tot} ^0$, which corresponds
to the fully spotted disk with zero $Q_{\rm tot}$. Next, we searched for the sky plane grid
pixels in succession that would give the maximum 
$Q_{\rm tot} ^{k+1}/F_{\rm tot} ^{k+1} =
(Q_{\rm tot} ^k + \Delta Q_{ij} ^{max}) / (F_{\rm tot} ^k + \Delta F_{ij} ^{max})$,
where $\Delta F_{ij} ^{max} = 4 \times (F_{ij} ^{\mathrm{phot}} - F_{ij} ^{\mathrm{spot}})$.
In parallel, the following condition was checked: 
$\Delta Q_{ij} ^{max} F_{\rm tot} ^k > \Delta F_{ij} ^{max} Q_{\rm tot} ^k$.
When the right part exceeded the left,
the remaining grid pixels were considered as occupied by spots, meaning that
their contribution to the total polarization due to a lower flux exceeded the
$\Delta Q_{ij} ^{max}$ contribution.

Figure \ref{fig_pmax} shows the highest polarization degree that can be reached by
different types of stars assuming the spot temperature contrasts from
\cite{2015MNRAS.448.3053A} for selected wavelengths corresponding to the effective
midpoints of $B$, $V,$ and $I$ broadband filters.
The solid lines mark the results obtained
with spherical atmosphere models, and the dashed lines show the plane-parallel models.
We note that spherical models include lower gravity log$g$ = 1, which is not available
for the plane-parallel model data. The two model sets match well
for higher
gravities and stars in the lower and upper effective temperature
ranges.
In general, the curves for the plane-parallel models appear to be less sensitive to the temperature,
especially for lower log$g$, while the spherical models show
a noticeable
depression in between 4500 and 5500 KK that grows at lower wavelengths.
This feature is discussed in detail in \cite{2017A&A...601A...6K}, who
considered the differences between plane-parallel and spherical models of a host star atmosphere
for exoplanet transit polarimetry. Similarly, the transit polarization
for spherical atmospheres in this temperature range was also found to be lower than for plane-parallel atmospheres.
This is explained by the noticeable temperature differences between atmospheric layers in the input
plane-parallel and spherical models, which affect the calculated center-to-limb intensity and
polarization profiles. Specifically, spherical models are systematically cooler than the
plane-parallel models by up to 500~K at $T_{\rm eff}$ = 5000~K
\citep[see Fig.4 in][]{2017A&A...601A...6K}, and this temperature
difference diminishes
rapidly toward higher $T_{\rm eff}$, similarly to what is observed in Figure \ref{fig_pmax}.

In Figure \ref{fig_pmaxdsk} we show several examples of disk asymmetries that result in the highest
linear polarization for the selected active stars. They show that the highest polarization
is obtained when most of a star is covered by spots, with the quiet photosphere confined
to narrow segments at the top and bottom in Fig.~\ref{fig_pmaxdsk}. For other stellar models we
found similar visible disk appearances with an intermediate ratio of spots and quiet photosphere.
Evidently, these asymmetries are not to be observed, and the
expected scattering polarization levels
due to the spots are probably significantly lower than the derived upper limits.

\subsection{Spot coverage dependency}

Using the scaling relations \eqref{eql1} and \eqref{eql2} for
a lognormal size distribution and
a random surface spot configuration, we next checked how the polarization depends on
the total spot coverage for different types of stars. As mentioned above,
the coverage was calculated as a fraction of the spherical grid elements
occupied by umbral and penumbral regions, as in \cite{2004MNRAS.348..307S}. According
to this definition, the polarization will not fully cancel out
and become zero for a 100\% spotted star because of the remaining
umbra-penumbra inhomogeneities. It can be noted that for a given size distribution and
total coverage, the random surface distribution statistically gives a lower limit
polarization constraint, and any systematic deviations from it will increase the disk
asymmetry and the net polarization.
                        
\begin{figure}
\includegraphics[]{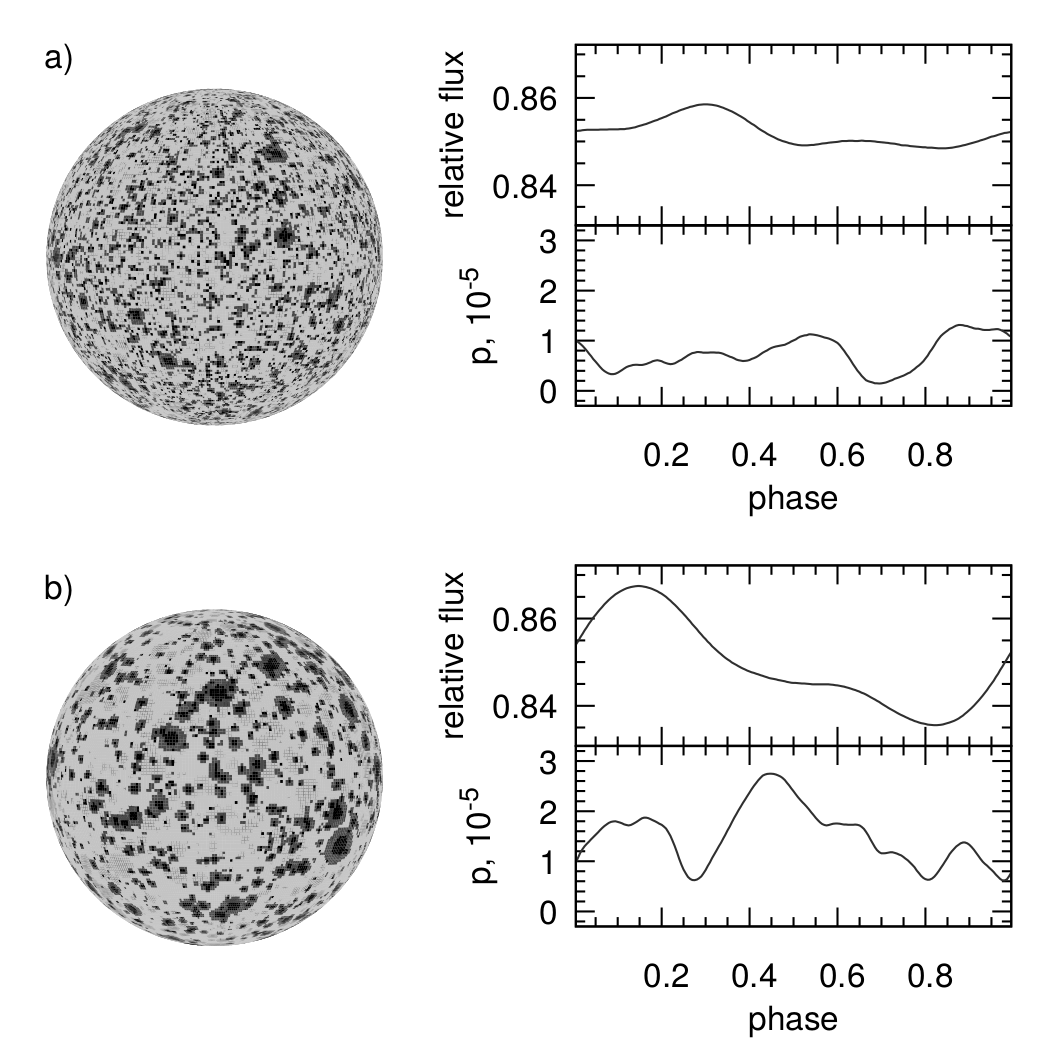}
\caption{Relative flux and polarization degree phase curves for a model M2IV star
with $T_{\rm eff}$ = 3500~K, $T_{\rm spot}$ = 3000~K, and log$g$ = 3.0 
at $\lambda$ = 4450 \text{\AA} for one
total spot coverage of 30\%, but different average spot sizes $\langle A \rangle$
(1.63 and 9.38 [$\times10^{-6} A_{1/2 *}$]), calculated using $n_A$ = 0.5 ($top$)
and 1.0 ($bottom$).
\label{fig_p_na}}
\end{figure}

\cite{2004MNRAS.348..307S} suggested that reasonable choices of the exponents ${n_A}$ and ${n_\sigma}$
in the scaling relations of Eqs. (\ref{eql1}), (\ref{eql2})
lie within the range of 0.5 to 1.0. Figure \ref{fig_p_na} illustrates how the choice of the scaling law for the average spot size affects
the polarization levels. The difference between linear and square-root laws is significant; it
amounts to almost a factor of 3 for the amplitude and the maximum polarization degree per
rotation period. Our tests for ${n_\sigma}$ showed a similar tendency with an even greater
discrepancy.

\begin{figure}
\includegraphics[]{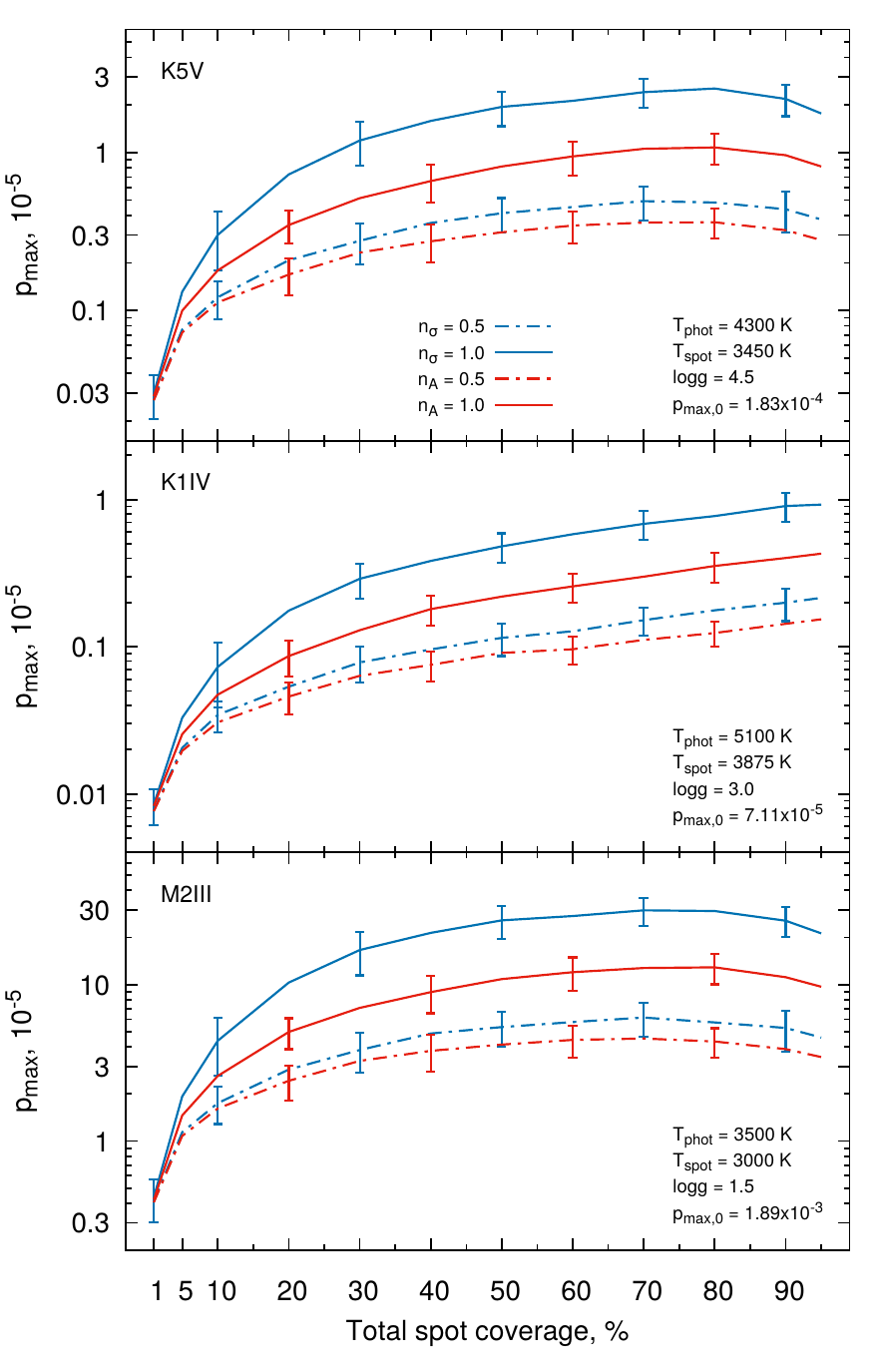}
\caption{Maximum polarization degree per rotation period for three stellar models
versus total spot coverage for $\rm \lambda$ = 4450 \text{\AA}. We show results for two
scaling laws for distribution width $\sigma_A$ (in blue) and average spot size
$\langle A \rangle$ (in red). The solid lines correspond to the linear
scaling law, and the dash-dotted lines show the square-root law.
\label{fig_p_ff}}
\end{figure}

To analyze the connection between the polarization and the total spot coverage, we
selected several model stars with different gravities and temperatures. The spot temperature
contrast for each model was found from Eq. \eqref{eq_dt}, and the stellar rotation axis was
directed parallel to the sky plane, meaning that it had an inclination of 90$^{\circ}$(see the next
subsection for the effect of inclination). Accounting
for the large polarization scatter obtained using different initial random seeds,
we ran 100 iterations for each spot coverage and then averaged the results.
In Figure \ref{fig_p_ff}
we plot the maximum polarization degree against the total spot coverage for
three model stars: 1) a K5V dwarf with $T_{\rm eff}$ = 4300~K, $T_{\rm spot}$ = 3450~K,
and log$g$ = 4.5, 2) a K1IV subdwarf with $T_{\rm eff}$ = 5100~K, $T_{\rm spot}$ = 3875~K,
and log$g$ = 3.0, and 3) an M2III giant with $T_{\rm eff}$ = 3500~K, $T_{\rm spot}$ = 3000~K,
and log$g$ = 1.5. In addition to considerable scatter, there is a large
difference between the lower and upper curves that is on the order of one magnitude.
We conclude from Figs. \ref{fig_p_na} and \ref{fig_p_ff} that the greater the
width of distribution $\sigma_A$ and average size $\langle A \rangle$, the more
inhomogeneous the produced surface maps, and consequently, the greater the disk asymmetries and
the higher the total polarization.
We also found that the average polarization degree values per rotation period
correlate with the maximum values, being in general twice as low, with
a small scatter between 1.8 and 2.2.
For selected model stars, the corresponding upper limit
constraints from the previous section are $\sim$10-100 times higher than the maximum
polarization degrees per rotation period, depending on the scaling law used.

No data are available on the ratio of umbra-to-penumbra area for
other stars, and it is questionable whether the solar value we
adopted here is universal and does not vary between different stellar types.
To check the influence of this parameter on our results,
we therefore ran several tests using the same parameters and scalings, but
generating spots without penumbra. As a result, we found very similar
correlations for polarization and spot coverage, within the scatter
shown in
Fig. \ref{fig_p_ff}.
This implies that the umbra-to-penumbra ratio plays a less important role in defining
polarization levels than the other parameters, such as size or surface
distributions.

\begin{figure}
\includegraphics[]{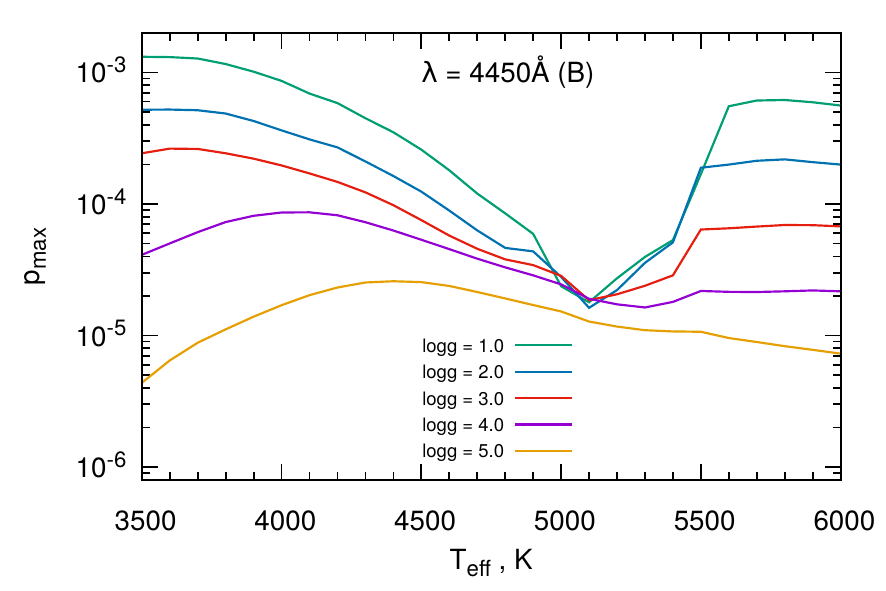}
\caption{Maximum polarization degree per rotation period for model stars with a spherical atmosphere, simulated for 30\% spot coverage and a lognormal spot size distribution
with a square-root scaling of the average spot area.
\label{fig_p_ff2}}
\end{figure}

For a total spot coverage of 30\%, the maximum
polarization degree per rotation period versus effective temperature for the stars with
different surface gravity, using spherical atmosphere models,
is shown in Fig. 5. For each model,
50 random spot configurations were simulated, and the results were averaged.
We used the scaling relation for the average spot area with ${n_A}$ equal to 0.5,
which gives a lower polarization (as seen in Fig. \ref{fig_pmax}). We also
checked the scaling for the distribution width with ${n_\sigma}$~=~1,
and it gives identical results over the whole temperature range, but higher
by a factor of 1.5. While tests with ${n_A}$~=~0.5
showed very small scatter with standard deviations of about 0.5\%,
for ${n_\sigma}$~=~1 the scatter was significantly larger, amounting to nearly 25\%.

Assuming that the selected two-temperature model, the photosphere-to-spot temperature contrasts,
and the spots size distributions are reasonable, the relations
shown in Figs. \ref{fig_p_ff}
and \ref{fig_p_ff2} present lower estimates of the linear scattering polarization for
a given set of stellar models. In the following we analyze how the different input
parameters influence our simulations.

\subsection{Inclination effect}

\begin{figure}
\includegraphics[]{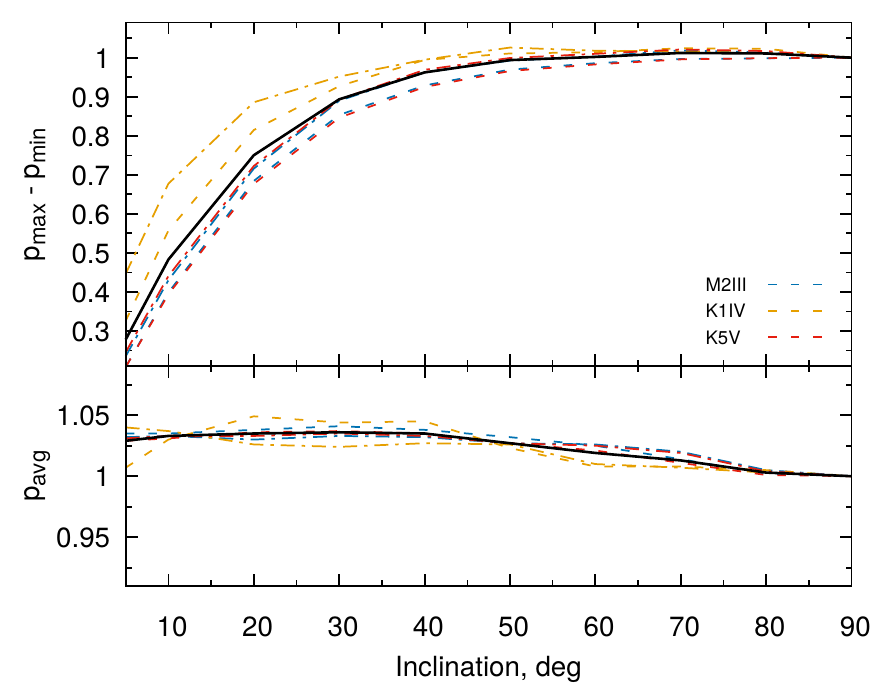}
\caption{Polarization degree amplitude ($top$) and average ($bottom$) values per
rotation period for different axis inclination angles at $\rm \lambda$ = $4450\rm{\AA}$,
(normalized to 90$^{\circ}$). Models
are the same as in Fig. \ref{fig_p_ff}.
For each model, different spot coverages were tested, using two scalings for $\sigma_A$ 
with ${n_\sigma}$, equal to 0.5 (dash-dotted line) and 1.0 (dashed line).
The average curve from all simulations is shown as the black solid line.
\label{fig_p_i}}
\end{figure}

Since our present model does not include spot evolution, the
Stokes parameters $Q$ and
$U$ for a star that is seen pole-on will vary harmonically during
the rotation period, producing a constant
polarization degree. Obviously, changing the inclination angle of the rotation axis from
0$^{\circ}$ (pole-on) to 90$^{\circ}$ will result in an increase of the polarization degree
amplitude. We selected the same stellar models as in the previous section and
generated 100 random spot configurations per spot coverage for each model, which ranged from 10\% to 90\%
with a step of 10\%, scaling the width parameter of the spot size distribution
using ${n_\sigma}$ = 0.5 and 1.0.
Each simulated spot configuration was used to calculate the polarization at all inclination
angles, the results were normalized to \textit{i} = 90$^{\circ}$ and then
averaged over 100 iterations. Analyzing the data, we did not find any dependency on the total spot
coverage and averaged results over this parameter.
Figure~\ref{fig_p_i} shows the normalized polarization degree amplitude and
its average over the rotation period versus the inclination of the rotation axis for the selected stellar
models, using two scaling laws. The gradual increase in average polarization toward
lower inclination angles is likely to result from the insufficient statistics and is not a
real feature, because the same curves as obtained using the median average showed
the decrease. When we analyzed the other dependencies, the K1IV model showed a noticeable deviation
from M2III and K5V, also for different scalings. In general, the average polarization
degree amplitude increases with the inclination of the rotation axis, rising more rapidly at
smaller inclinations and not changing significantly after \textit{i} > 40$^{\circ}$.

\subsection{Wavelength dependence}

\begin{figure}
\includegraphics[]{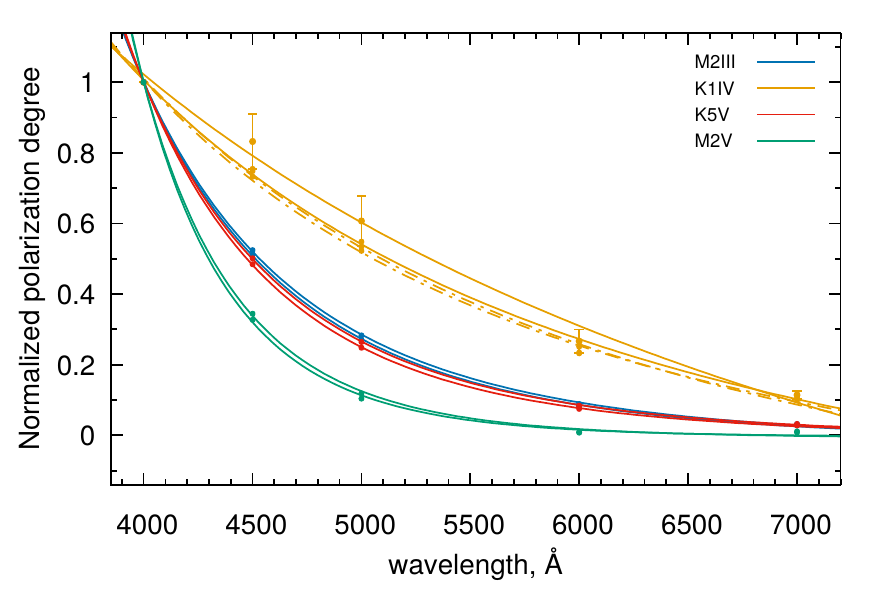}
\caption{Average polarization degree per rotation period at different wavelengths,
normalized to $\rm \lambda$ = $4000\rm{\AA}$. 
For each stellar model the overlying curves are the power-law best fit obtained
for 10\% total spot coverage, while the underlying curves are for 50\% coverage.
For K1IV, the results for two different size distributions are shown (see text).
\label{fig_p_wl}}
\end{figure}

To determine the wavelength dependence of the polarization, we used two
spot size distributions that resulted in lower and upper polarization estimates
for the stellar models in Fig. \ref{fig_p_ff}, scaling the average spot size using
${n_A}$ = 0.5 and the width of the distribution with ${n_\sigma}$ = 1.0.
They were used to generate surface maps with total spot coverages of 10\% and
50\%. Similarly to the inclination above, each simulated spot configuration
was run through all selected wavelengths. We performed 50 iterations per parameter set
and stellar model. The results were averaged and normalized to the wavelength of
$4000\rm{\AA}$. In Figure \ref{fig_p_wl} we show the dependencies for four different stellar models. Although Rayleigh scattering is the dominating process 
for the intrinsic polarization in most late-type stars, it is evident that
the wavelength dependencies show significant deviations from
the $\lambda^{-4}$
law, even though the input intrinsic polarizations for the quiet photosphere and spots separately
generally follow it. This is explained by the complex interplay of both limb-darkening and
polarization wavelength dependence for a given temperature of spots and
photosphere, when the total normalized Stokes parameters are calculated. Fits using the
power-law function for our entire model set resulted in a wide range of exponents
${n_\lambda}$ between -2 and -10.
We found that steeper curves are more characteristic for cooler stars with a higher surface
gravity. Tests also showed a minor dependence on the selected size distribution and total
spot coverage. Figure \ref{fig_p_wl} shows that curves
for 10\% and 50\% spot coverage nearly coincide for the K5V, M2III, and M2V models.
On the other hand, tests for the hotter K1IV model showed a higher degree of scattering for the different
iterations and a stronger dependency on
spot sizes and total coverage. To understand this, the following 
approximated ratio of polarization degrees $p$ at different wavelengths and distance $\mu$
from the disk center can be considered:

\begin{equation}
\frac{p^{\lambda_1}}{p^{\lambda_2}} \approx \\ 
\frac{I^{\lambda_1}_{\rm phot}~P^{\lambda_1}_{\rm phot} - I^{\lambda_1}_{\rm spot}~P^{\lambda_1}_{\rm spot}}{I^{\lambda_2}_{\rm phot}~P^{\lambda_2}_{\rm phot} - I^{\lambda_2}_{\rm spot}~P^{\lambda_2}_{\rm spot}}  \\
\frac{F^{\lambda_2}_{\rm tot}}{F^{\lambda_1}_{\rm tot}},
\end{equation}

\noindent where $P$ and $I$ are functions of $\mu$, as in Eqs. \eqref{eqlF}-\eqref{eqlU},
and $F_{\rm tot}$ is the total stellar flux, assuming no spots.
We found that the ratio is almost independent of $\mu$ for the
K5V, M2III, and M2V models, and it is close
to values in Fig. \ref{fig_p_wl}. This can explain the
convergence of results for these stellar models for different spot sizes and coverage. On the other
hand, for the K1IV model, the ratio was found to vary significantly with $\mu$, which
is visible in the wider variations in Fig. \ref{fig_p_wl}.

\subsection{Spot temperature contrast}

\begin{figure}
\includegraphics[]{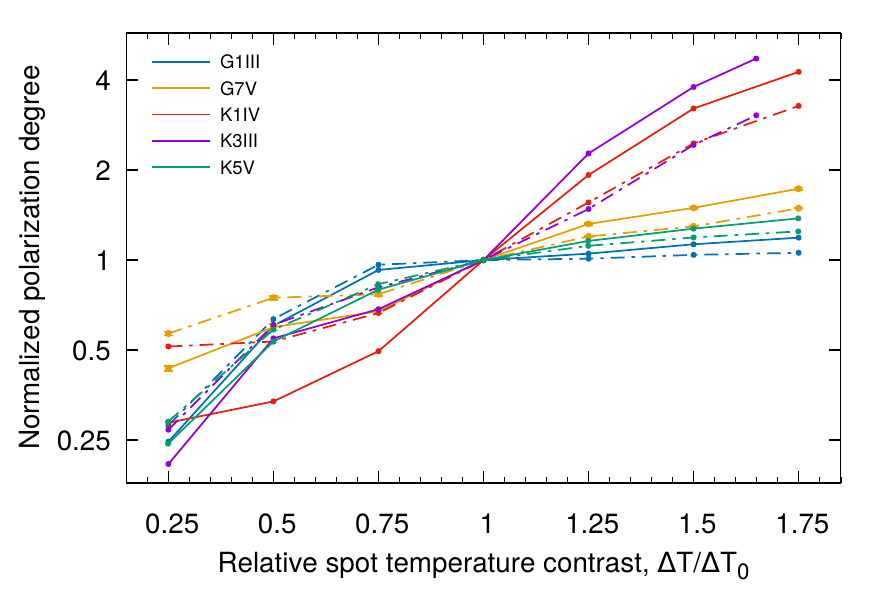}
\caption{Average polarization degree per rotation period versus spot temperature
contrast \citep[normalized to the contrasts in][]{2015MNRAS.448.3053A}.
For each stellar model, the dash-dotted lines show the 10\% total spot coverage, and the solid
lines show 50\% coverage.
\label{fig_p_ts}}
\end{figure}

Accounting for a small number of available measurements of spot temperatures for
different active stars, the \cite{2015MNRAS.448.3053A} linear fit that relates the spot contrast to
the effective temperature can be used only as a first approximation because the
real dependence is probably more complex and depends on more parameters.
We performed additional tests to verify how this parameter affects our
simulations. Like in previous subsections, the tests were made for several total spot coverages 
(scaling the width of the size distribution using ${n_\sigma}$ = 1.0)
with multiple runs averaged and normalized to values from Eq. \ref{eq_dt}.
Figure \ref{fig_p_ts} illustrates the average polarization degree per rotation period 
versus the relative spot temperature contrast for five different stellar models. 
All models showed a certain dependency on the total spot coverage and the choice of the
spot size distribution. As expected, the polarization grows with the spot
temperature contrast, with the average degree higher by 1.2 times to 4 times
in Fig. \ref{fig_p_ts} when the contrast is increased by a factor of 1.75.
Similar reverse trends are observed when the temperature contrast is decreased by
a factor of 4. In general, depending on selected stellar model, knowing the spot
temperature is important for linear polarization simulations.

\subsection{Active latitudes}

\begin{figure}
\includegraphics[width=8.9cm, angle=0]{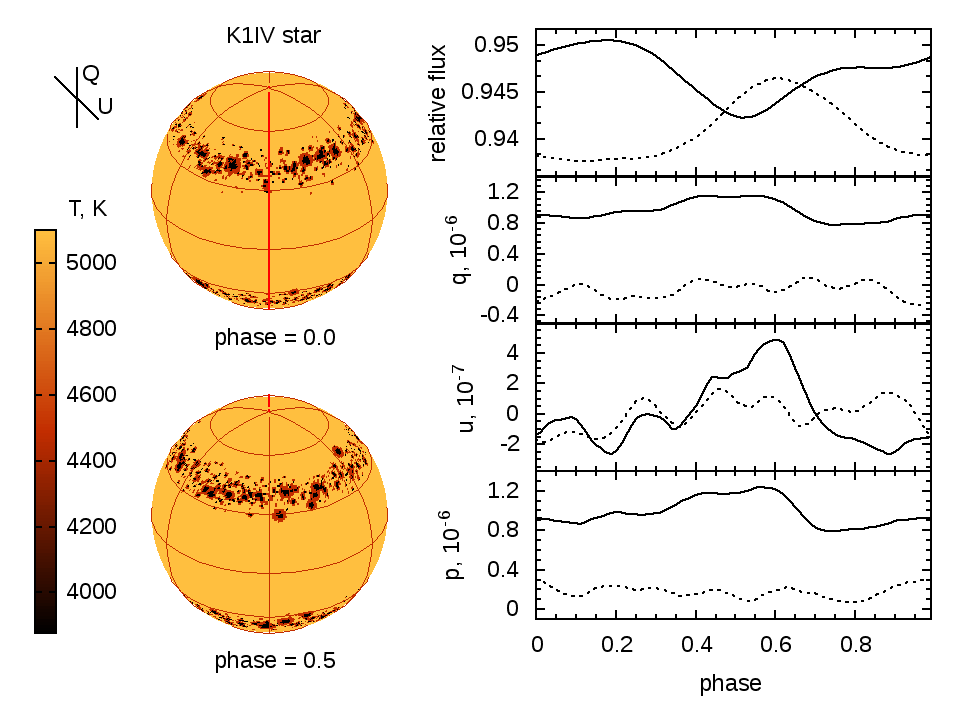}
\caption{Relative flux, normalized Stokes $Q$ and $U$ parameters, and polarization
degree, simulated for a K1IV star with active latitudes at
$\pm$40$^{\circ}$ and a total spot coverage of 10\% (solid lines). The same results for
a random surface spot distribution are shown with dashed lines.
\label{fig_lat}}
\end{figure}

As shown previously, there is a large scatter of polarization estimates 
for random spot configurations.
However, it can be shown that introducing any 
deviation from a random surface distribution for a fixed total spot coverage
is expected to systematically  increase the polarization degree.
Similarly to the Sun, where the sunspot occurrence is confined to
certain latitudes, we performed simulations for starspots by
selecting a Gaussian surface
distribution with different width over latitude in both hemispheres. Figure \ref{fig_lat}
shows our results of modeling active latitudes at $\pm$40$^{\circ}$ , with the width
equal to $\pm$10$^{\circ}$ for a K1IV star with 10\% total spot coverage.
One common feature that is also found below for a polar spot configuration is the upward
shift observed in the Stokes $Q$ phase curve, which increases the average polarization degree as well. For a given orientation in the sky plane, this can be explained schematically:
breaking the axisymmetry between the left and right parts of the disk results in residual total
Stokes $U$, while changing the axisymmetry between top right and bottom left parts affects
Stokes $Q$, as seen in this case. The Stokes parameters are
always defined in the instrumental system, therefore a comparison of our simulations and
observations requires knowing the stellar rotation axis orientation, which is rarely
known. However, with sufficient sensitivity, polarimetry is also a technique that
may allow determining the rotation axis orientation (e.g., from continuous phase curves of
recurring spots). That said, the degree of polarization
depends on the rotation axis inclination angle alone, being invariant to
the orientation of axis projection in the sky plane.

It is worth noting that only varying the width of the Gaussian distribution over
latitude for a fixed spot configuration can retain
the amplitudes of polarization parameters nearly unchanged.
However, depending on the inclination of the rotation axis combined with the location of active
latitudes, the phase curves can change significantly, which complicates the analysis.

\subsection{Polar spots}

Essentially, we found the same peculiarities in simulating polar spots as in
active latitudes: a shift in the Stokes parameters and an offset in
the degree of the polarization phase curves. Since the inclination angles of the rotation axis
of known active stars usually constitute 50-60 degrees or more, high-latitude
and polar spots can be seen closer to the limb than the lower active latitude spots, 
producing higher linear polarization. In addition,
reported large high-latitude spots, particularly in RS CVn systems, are often long-lasting
phenomena
\citep[e.g.,][]{2015MNRAS.447..567X, 2015A&A...578A.101K, 2015A&A...573A..98K, 2016arXiv160900196K},
which makes them suitable targets for limb-polarization effect detection, as discussed
on the example of XX Tri below.

\subsubsection{XX~Triangulum}

\begin{figure}
\includegraphics[width=8.9cm, angle=0]{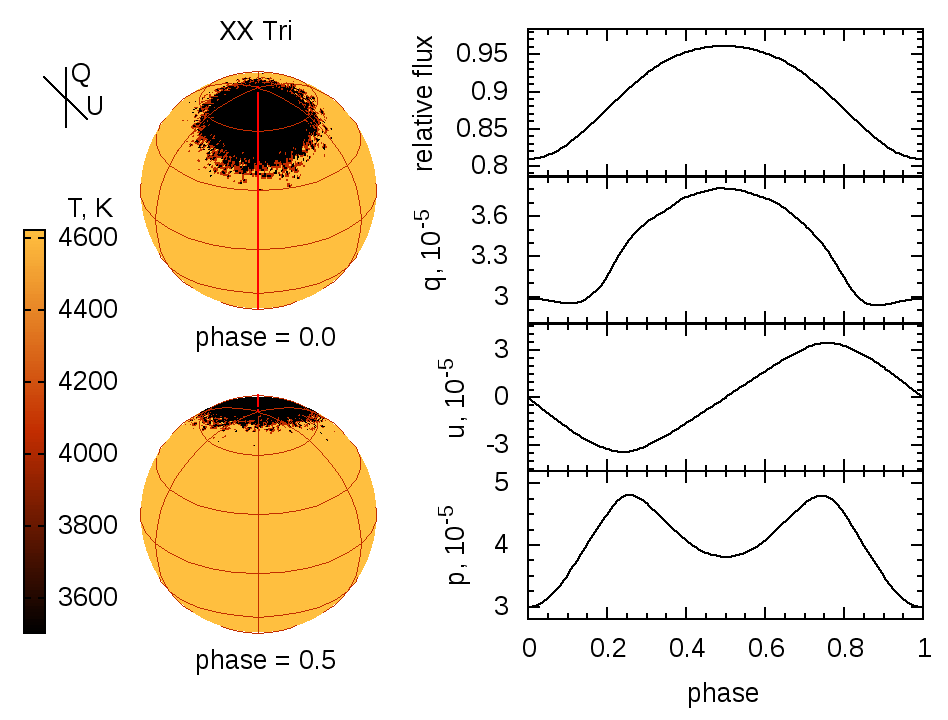}
\caption{Same as in Fig. \ref{fig_lat}, simulated for XX Tri with a polar spot configuration
and a total spot coverage of 8\%.
\label{fig_xxtri}}
\end{figure}

Polar and high-latitude spots have been reported for many stars,
based mainly on Doppler imaging  \citep[e.g., Table~2 in][]{2009A&ARv..17..251S},
and more recently also from interferometric observations, particularly for
$\lambda$~And \citep{2015arXiv150804755P} and $\zeta$ And \citep{2016Natur.533..217R}.
The spot sizes range from 0.1$\%$ to more than 10$\%$ of the stellar surface.
To estimate the linear polarization expected due to the disk asymmetry,
we selected the very active close RS CVn binary XX Tri (HD~12545), the bright component
of which is a spotted red giant of K0III spectral type. It is one of the record holders
in terms of the spot coverage, and is particularly known to have a polar superspot comprising about
10\% of the whole stellar surface, as shown by \cite{1999A&A...347..225S} from Doppler imaging.
In the later study by \cite{2015A&A...578A.101K}, conducted in a time span
of six years with the same technique, but using a different inversion code, the large
high-latitude spots were confirmed and their formation, merging, and decay were analyzed.

Taking the properties of XX~Tri as listed in \cite{2015A&A...578A.101K}, we simulated the
polar active region of the star with our code to estimate the linear polarization for it.
The active region was simulated as consisting of smaller spots using the scaling law
with ${n_A}$ = 1.0 and setting a Gaussian surface distribution centered on the latitude
of 70$^{\circ}$ (see Figure~\ref{fig_xxtri}). 
We adopted a total spot coverage from 5\% to 15\% and varied the Gaussian
width from 10$^{\circ}$ to 20$^{\circ}$, and found the resulting average polarization degree to lie between $3 \times 10^{-5}$ and $6 \times 10^{-5}$ with an amplitude range of
$1.2-1.8 \times 10^{-5}$ (the lower values were found for larger standard deviation).
Identical results are obtained when a single large spot of
the same size is added.
Apparently, a polarization of this degree can be measured even
with modern polarimeters such
as DIPOL-2 \citep{2014SPIE.9147E..8IP},
POLISH2 \citep{2015ApJ...800L...1W}, and HIPPI \citep{2015MNRAS.449.3064B}, which makes
XX~Tri a reasonable target for the limb-polarization effect detection.

\subsection{Active longitudes}

Active longitudes are a persistent surface feature that was reported for
many chromospherically active stars
\citep{1993A&A...278..449J, 1998A&A...336L..25B, 2001A&A...379L..30K}, including the Sun \citep{2003A&A...405.1121B}. The majority of
such stars are components of RS~CVn binaries or belong to single rapidly rotating
variables of FK~Com and BY~Dra type \citep{2005LRSP....2....8B, 2007IAUS..240..453K}.
Two long-lived active longitudes are often observed, separated by
$\sim$180$^{\circ}$, one showing a higher level of spot activity than the other.
The phenomenon of the switching between active longitudes is known as
flip-flop. For many stars, the flip-flop cycles of periods in the range of a few
years up to a decade have been revealed
\citep{2005LRSP....2....8B, 2007ApJ...659L.157B, 2007IAUS..240..453K}.
While being observationally well grounded, further theoretical studies are
needed to explain active longitudes and related flip-flop cycles
\citep[e.g.,][]{2005AN....326..278E, 2006A&A...445..703B}.

\begin{figure}
\includegraphics[width=8.9cm, angle=0]{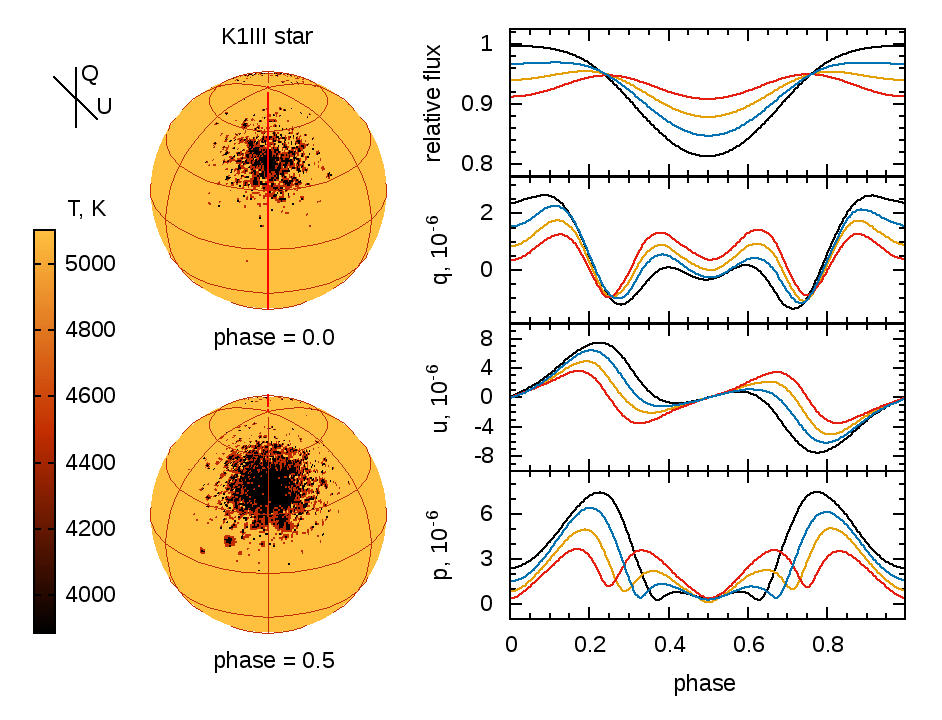}
\caption{Relative flux, normalized Stokes $Q$ and $U$ parameters, and polarization
degree, simulated for the K1III star (\textit{i} = 60$^{\circ}$) with two active longitudes
with different activity levels ratios: 0 (\textit{black}), 0.05 (\textit{blue}),
0.25 (\textit{yellow}), and 1.0 (\textit{red}).
The total spot coverage is 7\% for all cases.
\label{fig_alon}}
\end{figure}

In the framework of our modeling, we considered active longitudes in a simple
manner, for instance, excluding time evolution and differential stellar rotation. To
investigate characteristic polarization variation curves, we chose a model
K1III giant ($T_{\rm eff}$ = 5100~K, $T_{\rm spot}$ = 3885~K, log$g$ = 2.5),
which is common among the reported objects. The active regions were placed
on the opposite hemispheres at 45$^{\circ}$ latitude and 
simulated with the same parameters as for the XX~Tri polar region above.
Simulations were performed using different activity level ratios between the
two longitudes: 0.0 (only one active region), 0.05, 0.25, and 1.0 (equal-size regions).
The total spot coverage was fixed to 7\% for all cases. The calculation
results
are shown in Figure \ref{fig_alon}. Evidently, switching the dominant activity
to the second longitude shifts the polarization curves by 0.5 in phase.
In general, the curves for equally active longitudes revealed smaller amplitudes
with four similar peaks than the double-peaked more varied phase curves for one active longitude.
The results can also vary noticeably, depending
on the inclination and the latitudes of the active regions. At the same time, 
substituting the regions generated using a lognormal spot size distribution with single spots of
approximately the same size was found to affect the curves not
and only resulted in slightly
higher and narrower sinusoidal peaks.

\section{Conclusions}

We have theoretically investigated the scattering linear polarization effect
in continuum produced by starspots on active late-type stars. Assuming a simple two-temperature
photosphere model with circular non-evolving spots and using spot temperature contrasts
from \cite{2015MNRAS.448.3053A},
we developed a code to generate different surface spot configurations and calculate
the Stokes parameters and degree of polarization phase curves over the stellar rotation period.
As a first approximation, we used extrapolations from the solar to starspot size
distributions by \cite{2004MNRAS.348..307S} and selected random spot configurations. We
analyzed how different input parameters influence our simulations. Active
latitudes and longitudes and polar spots were then considered. The following conclusions were reached.

(i) The linear scattering polarization should be much higher in the blue spectral bands than in visual
and red bands. It is less dependent on the effective temperature, except in the region
between 4500 K and 5500 K, where the polarization drops significantly for the
spherical model atmospheres we used. The effect is more varied between luminosity classes,
with lower gravity stars showing higher polarization.

(ii) When we assumed a lognormal distribution of starspots by size, the polarization levels varied
significantly such that wider distributions and those with larger average spot sizes resulted in
generally higher polarization levels. When we fixed the total spot coverage, the difference between
the selected size distributions was found to be on the order of one magnitude.

(iii) The average polarization degrees for different stars increase gradually up to high spot
coverages of 70-80\%. However, the values obtained from different
random runs are spread significantly, such that it is hard to distinguish between selected spot size distributions.

(iv) Tests using random spot configurations showed a systematic increase in the amplitudes of
the polarization degree phase curves as the stellar rotation axis orientation changed from pole-on
to equator-on; it was steeper for axial inclination angles of less than 40$^{\circ}$.

(v) Despite a power-law wavelength dependence of the polarization, it was found to vary significantly
between different stellar models, which was explained by the interplay of wavelength
dependencies of the input limb-darkening and polarization for the two-temperature photosphere model we adopted.
Fitting the curves resulted in power-law exponents ranging from -2 to -10 for a given model set and
spot temperature contrasts. On the other hand, the wavelength dependences for most models
we obtained were almost not influenced by the choice of the spot size distribution and total coverage.

(iv) For a fixed spot coverage, any deviations from the random surface distribution are expected to increase
the average polarization of a star. Simulating reoccurring long-lasting formations, that is, active
latitudes and longitudes and polar caps, tended to increase the
average polarization degrees by
a certain constant background level.

(v) Because the axial inclination angles of known active stars are often more than
60$^{\circ}$, high-latitude and polar spots, which are closer to the limb, are expected to produce higher
and longer lasting polarization. Simulating the high-latitude spot configuration for XX Triangulum
using Doppler-imaging data, we estimated polarization degree for it to lie between $3 \times 10^{-5}$ and
$6 \times 10^{-5}$ with an amplitude range of $1.2-1.8 \times 10^{-5}$, which may be detected with modern polarimeters \citep[e.g.,][]{2016MNRAS.455.1607C}.

Overall, our simulations for late-type active stars show that the effect of linear scattering
polarization due to starspots is relatively small, with typical values of between
$10^{-6}$ and $10^{-4}$. However, these levels are characteristic for many problems of
upcoming astrophysical studies. While interesting on its own, the effect can complement
existing photometric and spectroscopic methods used in starspot studies.
In addition, the limb polarization effect is important for exoplanetary research
for planets orbiting active host stars, particularly for their atmosphere characterization,
where high-precision polarimetry will play an important role.

\begin{acknowledgements}
This work was supported by the European Research Council Advanced Grant HotMol (ERC-2011-AdG291659).
\end{acknowledgements}

\bibliographystyle{aa} 
\bibliography{aa30513-17.bib} 

\end{document}